\documentclass{article}
\usepackage{a4wide}
\usepackage{geometry}
\usepackage{caption}
\usepackage[english]{babel}
\usepackage{amssymb,mathtools} 	
\usepackage{amsfonts,amsthm}
\usepackage{newtxtext}       	
\usepackage{subeqnarray}	 	
\usepackage{tikz}
\usetikzlibrary{quantikz}

\usepackage{algorithm}
\usepackage{algpseudocode}
\usepackage{graphicx}        	
\usepackage{pict2e}
\usepackage[usenames,dvipsnames]{pstricks}
\usepackage{epsfig, epic, pst-all}
\DeclareGraphicsExtensions{.pdf,.eps, .png}
\graphicspath{ {./graphics/} }  
\usepackage{xcolor,colortbl}\theoremstyle{definition}

\usepackage{verbatim, float,xr-hyper}
\usepackage[colorlinks=true]{hyperref}
\newcommand{\op}[1]{\mathrm{#1}}						
\newcommand{\bcube}{\{0,1\}}							
\newcommand{\reellZ}{\mathbb{R}}						
\newcommand{\Landau}[1]{O{\left(#1\right)}}				
\newcommand{\bitplus}{\stackrel{{\scriptscriptstyle 2}}{\oplus}}		
\newcommand{\bitscal}{\stackrel{{\scriptscriptstyle 2}}{\odot}}		
\newcommand{\Hil}{\mathbb{H}}							
\newcommand{\qHil}{{\phantom{\Hil}}^{\scalebox{0.6}{q}}\!\Hil}		
\newcommand{\setsep}{\, \vert \;}							
\newcommand{\unOp}[1]{\mathbf{U}\!\left( #1 \right)}					
\newcommand{\solutionSet}{L}							
\newcommand{\function}[5]{\begin{array}{rccl}			
#1 : & #2 & \longrightarrow & #3 \\
& #4 & \longmapsto     & #5
\end{array} }
\newcommand{\I}{\mathrm{i}}								
\newcommand{\E}{\mathrm{e}}								
\newcommand{\betrag}[1]{\left\vert#1\right\vert}		
\newcommand{\one}{\mathbf{1}}							
\newcommand{\CNOT}{\textnormal{CX}}						
\DeclareMathOperator{\Mat}{Mat}							
\newcommand{\fadist}{\;\;}								
\newcommand{\norm}[1]{\left\vert\left\vert #1 \right\vert\right\vert}		
\newcommand{\card}[1]{{\vert #1 \vert}}					
\newcommand{\footremember}[2]{%
    \footnote{#2}
    \newcounter{#1}
    \setcounter{#1}{\value{footnote}}%
}
\newcommand{\footrecall}[1]{%
    \footnotemark[\value{#1}]%
} 
\begin{document}
\title{Oracle Operators for Non-Boolean Functions}
\date{July 14, 2022}
%
\author{
Fatema Elgebali\footremember{AIMS}{
AIMS Cameroon,
P.O. Box 608, Crystal Gardens, Limbe, Cameroon
}
\footnote{Cairo, Egypt, \url{fatema.elgebali@aims-cameroon.org}
}
\and
Wolfgang Scherer\footrecall{AIMS}
\footnote{Kingston, United Kingdom,
\url{wolfgang.m.scherer@gmail.com}
}
}
\maketitle
\abstract{We present a construction of a general oracle operator
for a real-valued function on the Boolean cube. As an
application, we use such operators in Shyamsundar's 
Non-Boolean Amplitude Amplification~\cite{art:Shyams2021} to solve binary 
optimization problems with a non-adiabatic algorithm.}
\vspace{2ex}

\maketitle

%
\section{Introduction}
\label{sect:Intro}
A large class of quantum algorithms makes use of oracular 
operators~\cite{internet:AlgorithmZoo}. One of the earliest of these
is Grover's search algorithm~\cite{inproc:Grover96a,art:Grover1997},
which uses an oracle operator based on a Boolean oracle function
to amplify the probability to find the items searched for.
The amplitude amplification method has since been generalized to
many other algorithms~\cite{internet:AlgorithmZoo,art:Brassard-et-al2002}
and recently Shyamsundar~\cite{art:Shyams2021} has proposed a further
generalization which works with real valued (rather than binary valued)
functions on the Boolean cube.

All operators required for the Non-Boolean amplitude amplification
can easily be implemented except for the conditional oracle operator 
$\op{U}_f$, for which hitherto no circuit for a general 
$f:\bcube^n \to \reellZ$ has been known.
In this paper we remedy this situation and present a
circuit for $\op{U}_f$ which is built from the Pauli-$\op{X}$, 
the phase multiplication and the conditional NOT operators.

Our construction makes use of the Fourier expansion on 
the Boolean cube~\cite{art:DeWolf2008}. In the most general case,
when the function requires all its Fourier coefficients for its
expansion, our construction is not efficient since we require a
sub-circuit for each Fourier coefficient. However, this is not 
the case for binary optimization problems, which encompass 
a large number of real world problems such as
Partitioning Problems, Binary Integer Linear Programming,
Coloring Problems and Traveling Salesman Problems~\cite{art:Lucas2014}.
For such problems we can directly calculate the Fourier coefficients
from the problem parameters without evaluating the function.
Moreover, the number of Fourier coefficients is $\Landau{n^d}$, where
$d$ is the degree of the binary polynomial, and $d = 2$ for all of
the aforementioned problems. 

Hence, we can use the Non-Boolean amplitude amplification 
proposed by Shyamsundar together with our construction 
of a general oracle operator to attempt solutions of the aforementioned
problems with a non-adiabatic algorithm.
As a method to solve an optimization problem in a non-adiabatic way, our 
algorithm is different from previously 
proposed methods which are based on variants of a
Grover search~\cite{art:Duerr-et-Hoyer1999,
art:Baritompa-et-al2005,art:Chakrabarty-et-al2017,art:Chen-et-al2019}.
Nevertheless, our circuit for a general oracle operator
may also be useful for these methods. 

This paper is organized as follows. In Section~\ref{sect:Not} we introduce
the necessary notation. In Section~\ref{sect:AA} we briefly review the
oracle operators used in Boolean and Non-Boolean amplitude amplifications
and outline which probabilities are amplified in the the Non-Boolean case
as well as how this can be used to find extrema of real valued functions.
In Section~\ref{sect:OracleOp} we present a circuit for a  
an oracle operator of a general real-valued
function on the Boolean cube $\bcube^n$. Since our construction is based
on the Fourier coefficients of the function, we start this section with
a brief review of the Fourier expansion on the Boolean cube. 
In Section~\ref{sect:BinOpt} we show how our construction of the general
oracle operator together with the Non-Boolean amplitude amplification
algorithm can be used to solve binary optimization problems. Finally,
Section~\ref{sect:Conc} offers some conclusions.
\section{Notation}
\label{sect:Not}
For any $x \in \{0, 1, \dotsc, 2^n-1 \}$ we denote its binary
expansion coefficients by $x_j \in \bcube$ such that
\begin{equation*}
x = \sum_{j=0}^n x_j 2^j
\, .
\end{equation*}
The binary expansion defines a bijection 
$b: \{0, 1, \dotsc, 2^n-1 \} \to \bcube^n$ and we refer
to $\bcube^n$ as the Boolean cube. 
With the help of this bijection, any function 
$h: \{0, 1, \dotsc, 2^n-1 \} \to \reellZ$ defines a function
$f = h \circ b : \bcube^n \to \reellZ$ and vice versa. Hence,
any construction for the real vector space of all real-valued
functions on the Boolean cube can also be applied to the
real vector space of all real-valued functions on
$\{0, 1, \dotsc, 2^n-1 \}$ and we will thus only work with
functions $f: \bcube^n \to \reellZ$.

For $u,v \in \bcube$ we denote their binary addition by
\begin{equation*}
u \bitplus v : = (u + v) \mod 2 
\, ,
\end{equation*}
such that $\bitplus : \bcube \times \bcube \to \bcube$.

For any bit-strings $a, b \in \bcube^n$ we define the modular bit-product of $a$ and $b$ as
\begin{equation*}
a  \bitscal b : = \Bigl(\sum_{j=0}^{n-1} a_j b_j\Bigr) \mod 2
\, ,
\end{equation*}
such that $ \bitscal : \bcube^n \times \bcube^n \to \bcube$.

We set $[n]:= \{ 0, \dotsc, n-1 \}$ and let $P[n]$ denote 
the power set of $[n]$, that is the set of all subsets of $[n]$.

We denote the two-dimensional qubit Hilbert space by $\qHil$ and 
the computational basis states in its $n$-th tensor power $\qHil^{\otimes n}$ 
by $\ket{x}$, where
\begin{equation*}
\ket{x} 
=
\ket{x_{n-1}} \otimes \dotsm \otimes \ket{x_0}
\in 
\qHil \otimes \dotsm \otimes \qHil
= 
\qHil^{\otimes n}
\, .
\end{equation*}
Our input/output or work register will be $\Hil_W = \qHil^{\otimes n}$, and
we will augment this with the ancilla register $\Hil_A = \qHil$ to form
the total quantum register $\Hil_W \otimes \Hil_A$. The computational basis
in $\Hil_W \otimes \Hil_A$ will then be comprised of vectors $\ket{x} \otimes \ket{a}$
where $x \in \bcube^n$ and $a \in \bcube$.

We denote the group of unitary operators on a Hilbert space $\Hil$ by
$\unOp{\Hil}$.
%
\section{Amplitude Amplification}
\label{sect:AA}
\subsection{Boolean Amplitude Amplification}
\label{subsect:AA:BAA}
Boolean amplitude amplification is a method used in a variety of quantum algorithms.
It helps to solve the following problem: Suppose we are given a problem
that has a set $\solutionSet \subset \bcube^n$ of solutions which 
can be encoded in vectors in a subspace $\Hil_{\solutionSet}$
of the input/output Hilbert space $\Hil_W = \qHil^{\otimes n}$ of our quantum system. In other words,
if we find a vector in $\Hil_{\solutionSet}$, we have solved the problem.

To accomplish this, we initialize our quantum system in an initial state that we can
easily prepare. Boolean amplitude amplification is then a method of rotating such an initial state
into the solution subspace $\Hil_{\solutionSet}$. The rotation into the solution subspace is done 
by repeatedly applying a number of unitary operators for an optimal number of iterations.

One of these unitary operators is the so-called oracle operator, which helps to identify the
solution subspace. It is one of the main ingredients of the Boolean amplitude amplification  algorithm and 
makes use of a Boolean oracle function which helps to identify the elements of the solution set.
Formally we define a Boolean oracle function $g$ as
\begin{equation*}
\function{g}{\bcube^n}{\bcube}{x}{
g(x) = \begin{cases}
			0 \text{ if } x \not\in \solutionSet \\
			1 \text{ if } x \in \solutionSet \, .
		\end{cases}
}
\end{equation*}
The Boolean oracle operator $\op{U}^{(b)}_g$ is a unitary operator
on $\Hil_W \otimes \Hil_A$. It is defined by its action on the computational 
basis $\ket{x} \otimes \ket{a}$ in $\Hil_W \otimes \Hil_A$ by
\begin{equation*}
\op{U}^{(b)}_g \ket{x} \otimes \ket{a}
= 
\ket{x} \otimes \ket{a \bitplus g(x)} 
\, ,
\end{equation*}
and then by linear continuation on all of $\Hil_W \otimes \Hil_A$. A consequence of
this definition is that, 
with $\ket{-} = \frac{\ket{0}-\ket{1}}{\sqrt{2}}$, 
the oracle $\op{U}^{(b)}_g$ acts on the state $\ket{x} \otimes \ket{-}$ as follows
\begin{equation*}
\op{U}^{(b)}_g \ket{x} \otimes \ket{-}
=
(-1)^{g(x)} \ket{x} \otimes \ket{-}
= 
\E^{\I\pi g(x)} \ket{x} \otimes \ket{-}
\, ,
\end{equation*}
and it is this multiplication with $-1$ of states $\ket{x} \otimes \ket{-}$ for which
$x \in \solutionSet$ that makes $\op{U}^{(b)}_g$ a useful ingredient in the Boolean
amplitude amplification.
%
\subsection{Non-Boolean Amplitude Amplification}
\label{subsect:AA:NBAA}
Recently a generalization of the Boolean amplitude amplification has been
proposed by Shyamsundar~\cite{art:Shyams2021} and in this section we briefly summarize 
this so-called Non-Boolean amplitude amplification algorithm.
In doing so, we continue to use our order $\Hil_W \otimes \Hil_A$ of
work and ancilla register, which is the reverse of the one used in~\cite{art:Shyams2021}.

Whereas the original Boolean amplitude amplification worked with a Boolean
oracle function $g: \bcube^n \to \bcube$, the Non-Boolean generalization
introduced by Shyamsundar works for any real-valued function
\begin{equation*}
f: \bcube^n \to \mathbb{R}
\, .
\end{equation*}
The crucial ingredient of the Non-Boolean amplitude amplification
is the (conditional) oracle operator $\op{U}_f \in \unOp{\Hil_W \otimes \Hil_A}$ which is defined as
\begin{equation}
\label{eqn:AA:NBAA.20}
\op{U}_f
=
\sum_{x \in \bcube^n} 
\ket{x}\bra{x} \otimes \bigl(\E^{\I f(x)} \ket{0}\bra{0}   + \E^{-\I f(x)}\ket{1}\bra{1} \bigr) 
\, .
\end{equation}
The conditional oracle operator performs arbitrary phase-shifts $f(x)$ 
on the computational basis states $\ket{x}$ conditional on the ancilla state.
It is this conditional oracle operator for which we shall present a quantum circuit
in Section~\ref{sect:OracleOp}.

In the Non-Boolean amplitude amplification algorithm we initialize the ancilla in the state
$\ket{+} = \frac{\ket{0} + \ket{1}}{\sqrt{2}} $ and define the initial total-register state 
as
\begin{equation}
\label{equa:AA.ExtremNBAA.30}
\ket{\Psi_0}
=
 \ket{\psi_0} \otimes \ket{+}
\in \Hil_W \otimes \Hil_A
\, ,
\end{equation}
where
\begin{equation*}
\ket{\psi_0}
=
\sum_{x \in \bcube^n} a_0(x) \ket{x} \in \Hil_W
\end{equation*}
is the initial state satisfying $\norm{\psi_0} = 1$
in the work register. 
Typically, $a_0(x) = 1/2^{n/2}$, 
and in this case
$\ket{\Psi_0}$ can easily be prepared by the application of the 
appropriate tensor power of the Hadamard operator on
$\ket{0}^{n+1} \in \Hil_W \otimes \Hil_A$, that is
\begin{equation}
\label{equa:AA.ExtremNBAA.45}
\ket{\Psi_0}
=
\op{H}^{\otimes n+1} \ket{0}^{n+1}
\, .
\end{equation}
For a successful execution of the Non-Boolean amplitude amplification algorithm
we need the parameter $\theta \in [0,\pi]$ that is implicitly defined by
\begin{equation*}
\cos(\theta)
=
\sum_{x \in \bcube^n} p_0(x) \cos\bigl( f(x) \bigr)
\, ,
\end{equation*}
where $p_0(x) = \betrag{a_{0}(x)}^2$ is the probability to find $x$
when measuring the initial state $\ket{\psi_0}$ in
the work register $\Hil_W$ in the computational basis.
As shown in \cite{art:Shyams2021}, the determination of $\theta$
can be achieved by making use of the quantum phase estimation algorithm \cite{art:Svore-et-al2014}.

The algorithm also requires the operator
\begin{equation*}
\op{S}_{\Psi_0}
= 
2 \ket{\Psi_0}\bra{\Psi_0} - \one^{\otimes n+1}
\in \unOp{\Hil_W \otimes \Hil_A}
\, ,
\end{equation*}
which amounts to a reflection about $\ket{\Psi_0}$ and can also be easily
implemented by making use of \eqref{equa:AA.ExtremNBAA.45}.

The Non-Boolean amplitude amplification is an iterative algorithm and 
is summarized as Algorithm~\ref{alg:SumNBAA}.
\begin{algorithm}[H]
\caption{Summary of Non-Boolean Amplitude Amplification Algorithm~\cite{art:Shyams2021}}
\label{alg:SumNBAA}
\begin{algorithmic}
\State 
 the system in the initial state 
$\ket{\Psi_0} \in \Hil_W \otimes \Hil_A$ as in \eqref{equa:AA.ExtremNBAA.30}
\While{$j \leq K$}
\If{$j$ is even} 
\State
Set $\ket{\Psi_{j+1}} = \op{S}_{\Psi_{0}} \op{U}^\dagger_f \ket{\Psi_{j}}$ 
\ElsIf{$j$ is odd}
\State
Set $\ket{\Psi_{j+1}} = \op{S}_{\Psi_{0}} \op{U}_f \ket{\Psi_{j}}$ 
\EndIf
\State $j \gets j+1$
\EndWhile
\State
Measure the ancilla register $\Hil_A$ in the basis $\{\ket{0}, \ket{1}\}$
\State
Measure the work register $\Hil_W$ in the computational basis 
$\bigr\{\ket{x} \setsep x \in \bcube^n \bigr\}$
\end{algorithmic}
\end{algorithm}

The iterations amplify the basis states amplitudes. 
The measurement of the ancilla at the end of the algorithm results 
with equal probability in either $\ket{0}$ or $\ket{1}$. 
It is performed to ensure that the ancilla register $\Hil_A$ and the 
work register $\Hil_W$ are not entangled in the final state of the algorithm.

Let $p_{K}(x)$ denote the probability to find $x \in \bcube^n$ 
when measuring $\Hil_W$ in the computational basis after $K$ iterations. 
The change when compared to the initial probability $p_{0}(x)$ 
is given by
\begin{equation}
\label{equa:AA.ExtremNBAA.60}
\frac{p_K(x) - p_0(x)}{p_0(x)}
=
\lambda_K(\theta) \Bigl( \cos(\theta) - \cos\bigl(f(x)\bigr) \Bigr)
\, ,
\end{equation}
where
\begin{equation}
\label{equa:AA.ExtremNBAA.70}
\lambda_K(\theta) 
=
\dfrac{\cos(\theta) - \cos((2K + 1) \theta)}{\sin^2(\theta)}
\, .
\end{equation}
From \eqref{equa:AA.ExtremNBAA.70}, we see that $\lambda_K(\theta)$ is an oscillatory function
of $K$ that has an upper bound, 
which is the optimal value of $\lambda_K(\theta)$, that is
\begin{equation*}
\lambda_K(\theta) \leq \lambda_{\text{opt}}(\theta)
=
\frac{1}{1 - \cos(\theta)}
\end{equation*}
The number of iterations $K$ is determined by the $\lambda_K(\theta)$ that is good enough to 
amplify the basis states with lower values of $\cos(f(x))$. 
From \eqref{equa:AA.ExtremNBAA.70} we see that $\lambda_K(\theta)$ 
increases monotonically for all $K$ such that
\[0 \leq K \leq \Big\lfloor \frac{\pi}{2\theta} \Big\rfloor. \]
Hence, the number of iterations that maximizes $\lambda_K(\theta)$ is determined as
\begin{equation*}
\tilde{K}
=
\Big\lfloor \frac{\pi}{2\theta} \Big\rfloor
\, .
\end{equation*}
Thus, we see from \eqref{equa:AA.ExtremNBAA.60} that,
if we iterate $\tilde{K}$ times such that $\lambda_{\tilde{K}} > 0$, 
then the Non-Boolean amplitude amplification provides a method
to amplify the probability to find the elements $x \in \bcube^n$ 
for which $\cos(f(x))$ is smaller than $\cos(\theta)$.
Moreover, the bigger  $\cos(\theta) - \cos(f(x))$, the bigger the 
amplification.

Alternatively, if one iterates to a $K$ such that
$\lambda_{K}(\theta) < 0$, then the probability to find 
those $x \in \bcube^n$ for which $\cos(f(x))$ is bigger than
$\cos(\theta)$ is amplified, and,
the bigger  $\cos(f(x))- \cos(\theta)$, the bigger the 
amplification in this case. 
%
\subsection{Finding Extrema with Non-Boolean Amplitude Amplification}
\label{subsect:AA.ExtremNBAA}
These properties of the Non-Boolean amplitude amplification
can be utilised to find the extrema of any function $F$ on 
the Boolean cube as follows.
Since $\bcube^n$ is a finite set, any function $F$ on it is bounded.
Let the bounds be $f_{-} \leq F(x) \leq f_{+}$. Then we have
\begin{equation}
\label{equa:AA.ExtremNBAA.105}
f_{\pm}(x) 
:= \
\pm \frac{f_{\pm} - F(x)}{f_{+} - f_{-}} \frac{\pi}{2}
\in [0, \frac{\pi}{2}]
\, .
\end{equation}
To show how we can use the Non-Boolean amplitude amplification to find the extrema
of $F$, we introduce the following notation.
\begin{eqnarray*}
F_{\min/\max}
& := &
\min/\max\{F(x) \setsep x \in \bcube^n \} 		
\\
\{x\}_{\min/\max}
& := &
\{x \in \bcube^n \setsep F(x) = F_{\min/max} \} 
\\
f_{\pm, \min/\max}
& := &
\min/\max\{f_{\pm}(x) \setsep x \in \bcube^n \} 		
\\
\{x\}_{\pm, \min/\max}
& := &
\{x \in \bcube^n \setsep f_{\pm}(x) = f_{\pm, \min/max} \} 
\, .
\end{eqnarray*}
The extrema of the $f_{\pm}$ coincide with those of $F$, that is, we have
\begin{equation}
\label{equa:AA.ExtremNBAA.110}
\begin{array}{rcl}
\{x\}_{+, \min}
& = 
\{x\}_{\max}
& = 
\{x\}_{-,\max} 
\\
\{x\}_{+, \max}
& = 
\{x\}_{\min}
& = 
\{x\}_{-,\min} 
\, .
\end{array}
\end{equation}
Hence, if we apply the Non-Boolean amplitude amplification to
$f_{-}$ for $\tilde{K}_{-}$ times such that $\lambda_{\tilde{K}_{-}}(\theta_{-}) > 0$,
then the $x$ for which $f_{-}(x)$ is maximal will experience the biggest
probability amplification. In other words, the probability to find
an $x \in \{x\}_{\max}$ will be amplified the most.
 
Likewise, if we apply the Non-Boolean amplitude amplification to
$f_{+}$ for $\tilde{K}_{+}$ times such that $\lambda_{\tilde{K}_{+}}(\theta_{+}) > 0$,
then the $x$ for which $f_{+}(x)$ is maximal will experience the biggest
probability amplification, that is all elements in $\{x\}_{\min}$ will experience
the largest probability enhancement. 

The scaling of $\pi/2$ in \eqref{equa:AA.ExtremNBAA.105} can be
replaced by $\pi/4$ (or any other value in $[0,\pi/2]$).
Reducing the scaling will decrease the width of the range of $f_{\pm}$
and the value of $\theta_{\pm}$ in their respective algorithms.
While this will increase the value of the optimal amplification
factors $\lambda_{\tilde{K}_{\pm}}(\theta_{\pm})$, 
numerical evidence suggests that it does not result
in greater amplification since 
$\cos(\theta_{\pm})-\cos\bigl( f_{\pm}(x) \bigr)$ shrinks in equal
measure.

In Section \ref{subsect:BinOpt.QuadBO} we apply this method to a simple
toy example from quadratic binary optimization.
%
\section{Constructing the Non-Boolean Oracle Operator}
\label{sect:OracleOp}
\subsection{Fourier Coefficients}
\label{subsect::OracleOp.FourierCoeff}
Our construction of $\op{U}_f$ makes use of the Fourier expansion of
functions on the Boolean cube $\bcube^n$. Hence, we start this section with
a brief review of this Fourier expansion, which is based on \cite{art:DeWolf2008}.

The set
\begin{equation*}
B_n := \{ f: \{0, 1\}^n \to \reellZ \}
\end{equation*}
forms a $2^n$-dimensional real vector space. We define
the inner product of two functions $f, g \in B_n$ as
\begin{equation}
\label{equa::OracleOp.FourierCoeff.20}
\langle f, g \rangle = \frac{1}{2^n} \sum_{x \in \{0, 1\}^n} f(x) g(x)
\, .
\end{equation}
We can identify each $S \subset [n] := \{0, 1, \ldots, n - 1\}$ 
uniquely by a bit-string of length $n$ denoted by
\begin{equation*}
\hat{S} = (s_{n-1}, s_{n-2}, \ldots, s_0)
\, ,
\end{equation*}
where
\begin{equation*}
	s_j = \begin{cases}
		0 &\text{ if } j \notin S \\
		1 &\text{ if } j \in S.
	\end{cases}
\end{equation*}
For any $S \subseteq [n]$, we define the parity function $\chi_S$ as
\begin{equation}
\label{equa::OracleOp.FourierCoeff.50}
\function{\chi_S}{\bcube^n}{\{\pm 1 \}}{x}{(-1)^{\hat{S}  \bitscal x}}
\, .
\end{equation}
For example, the parity functions of the sets $S$ with cardinality $\betrag{S} \leq 2$ are
\begin{equation}
\label{equa::OracleOp.FourierCoeff.55}
\begin{aligned}
\chi_{\emptyset}(x) & =  1 \\
\chi_{\{i\}}(x)		& =  (-1)^{x_i} = 1 - 2x_i \\
\chi_{\{i, j\}}(x)	& =  (-1)^{x_i + x_j} = (1 - 2x_i)(1 - 2x_j) = 1 - 2x_i - 2x_j + 4x_ix_j.
\end{aligned}
\end{equation}
Incidentally, the identification of each $S \subset [n]$ with its bits-string $\hat{S} \in \bcube^n$
also exhibits the cardinality of the power set $P[n]$ as $\card{P[n]} = 2^n$, and
the set $\{ \chi_S \setsep S \in P[n] \}$ forms an orthonormal basis (also called Fourier basis) 
of the $2^n$-dimensional real vector space $B_n$. 

For any function $f: \{0, 1\}^n \to \mathbb{R}$, we define a function
\begin{equation*}
\function{\hat{f}}{P[n]}{\reellZ}{S}{\langle f, \chi_S \rangle}
\, .
\end{equation*}
The real number $\hat{f}(S)$ is called the Fourier coefficient of $f$ at $S$ and 
the set of the Fourier coefficients is called the Fourier spectrum of $f$.
The linear map $\hat{}: \{f: \{0, 1\}^n \to \mathbb{R}\} \to \{\hat{f}: P[n] \to \mathbb{R}\}$ 
is called the Fourier transform of $f$.

It can be shown that the Fourier coefficients of a function $f \in B_n$
are its expansion coefficients in the basis $\{ \chi_S \setsep S \in P[n] \}$, that is,
that we have
\begin{equation*}
f(x) = \sum_{S\in P[n]} \hat{f}(S)\chi_S(x)
\end{equation*}
and thus
\begin{equation}
\label{equa::OracleOp.FourierCoeff.80}
\E^{\I f(x)} 
=
\prod_{S \in P[n]} \E^{\I \hat{f}(S)\chi_S(x)}
\, .
\end{equation}
We will use this fact to construct a circuit for $\op{U}_f$ as follows.
Suppose for each $S \subset [n]$ and $\alpha \in \reellZ$
we have a circuit to implement the operator
\begin{equation}
\label{equa::OracleOp.FourierCoeff.90}
\op{U}_{S}(\alpha)
=
\sum_{x \in \bcube^n} 
\ket{x}\bra{x}
\otimes
\bigl(
\E^{\I\alpha\chi_S(x)} \ket{0}\bra{0}
+
\E^{-\I\alpha\chi_S(x)} \ket{1}\bra{1} 
\bigr) 
\, .
\end{equation}
Then it is easy to see that, for any $S_1, S_2 \subset [n]$, 
the operators $\op{U}_{S_1}(\alpha)$ and $\op{U}_{S_2}(\alpha)$ commute,
and with \eqref{equa::OracleOp.FourierCoeff.80} that
\begin{equation*}
\prod_{S\in P[n]} \op{U}_S \bigl( \hat{f}(S)\bigr)
=
\sum_{x \in \bcube^n} 
\ket{x}\bra{x}
\otimes
\bigl(
\E^{\I f(x)} \ket{0}\bra{0}
+
\E^{-\I f(x)} \ket{1}\bra{1} 
\bigr) 
\, .
\end{equation*}
Hence, it follows from \eqref{eqn:AA:NBAA.20} that
\begin{equation}
\label{equa::OracleOp.FourierCoeff.110}
\prod_{S\in P[n]} \op{U}_S \bigl( \hat{f}(S)\bigr)
=
\op{U}_f
\, ,
\end{equation}
and if we can implement $\op{U}_S$, then we also have a circuit for $\op{U}_f$.

Before we turn our attention to constructing a circuit for $\op{U}_S$, it is worth pointing out
that in cases where $f$ is such that in its Fourier expansion the majority of its
$2^n$ Fourier coefficients $\hat{f}$ do not vanish,
the construction of $\op{U}_f$ given in \eqref{equa::OracleOp.FourierCoeff.110} 
involves $\Landau{2^n}$ sub-circuits of the type $\op{U}_S$.
Hence, this construction is not efficient in general. 

However,
there is a large number of instances, including NP-hard problems~\cite{art:Lucas2014}, where the number of 
non-zero Fourier coefficients is actually low and the non-zero coefficients can be easily obtained. 
In such cases our construction 
enables the application of the Non-Boolean amplitude amplification method to attack
these problems. For example, as we will show in Section~\ref{sect:BinOpt}, 
for binary optimization problems of functions which
are polynomial of degree $m$, we not only have just $\Landau{n^m}$ non-zero 
Fourier coefficients, but we can also easily calculate them from the
coefficients in the polynomial without the need to 
obtain $\langle f, \chi_S \rangle$ by explicitly calculating the right
side of \eqref{equa::OracleOp.FourierCoeff.20}.
%
\subsection{Implementation of $\op{U}_S$}
\label{subsect:OracleOp.ImplU_S}
Since $\hat{S} \bitscal x \in \{0,1\}$, it follows from \eqref{equa::OracleOp.FourierCoeff.50} that
\begin{equation}
\label{equa:OracleOp.ImplU_S.10}
\chi_S(x)
=
(-1)^{\hat{S} \bitscal x}
=
1 - 2\hat{S} \bitscal x
\, .
\end{equation}
For our construction of $\op{U}_S$ we will thus first 
construct a circuit for an operator $\op{U}_{\hat{S}\bitscal} \in \unOp{ \Hil_W \otimes \Hil_A}$ 
that acts on the computational basis $\ket{x} \otimes \ket{a}$ as
\begin{equation}
\label{equa:OracleOp.ImplU_S.20}
\op{U}_{\hat{S} \bitscal } \ket{x} \otimes \ket{a}
=
\ket{x} \otimes \ket{a \bitplus (\hat{S} \bitscal x)}
\, . 
\end{equation}
This operator will then be used to write $\hat{S} \bitscal x$ into the ancilla $\Hil_A$,
and then we will use conditional multiplication operators to apply phase shifts by the Fourier
coefficients $\hat{f}(S)$. This will result in $\op{U}_S$ as specified in 
\eqref{equa::OracleOp.FourierCoeff.90}.

To construct the circuit $\op{U}_{\hat{S}\bitscal}$, 
let $\CNOT_n(c, t) \in \unOp{\Hil_W}$ be the controlled-NOT operator, 
where $c$ is the control qubit and $t$ is the target qubit.
We can express the action of $\CNOT_n(c, t)$ on a computational basis
state $\ket{x} \in \unOp{\Hil_W}$ as
\begin{equation*}
\CNOT_n(c, t) \ket{x}
=
\ket{x_{n-1}} \otimes \dotsm \otimes \ket{x_{t+1}} 
\otimes \ket{x_c \bitplus x_t} \otimes
\ket{x_{t-1}} \otimes \dotsm \otimes \ket{x_{0}} 
\end{equation*}
Next, let $S \subset [n]$ be such that $S = \{j_1, j_2, \ldots, j_{\card{S}}\}$,  
where $j_1 < j_2 < \ldots < j_{\card{S}}$. 
With the controlled-NOT we define two operators $\op{R}_{S}$ and $\op{V}_{S}$.
The operator $\op{R}_{S}$ acting on 
$\Hil_W \otimes \Hil_A$ is defined as
\begin{eqnarray*}
\op{R}_S
& = &
\begin{cases}
\prod_{l = 1}^{\card{S} - 1} \CNOT_n( j_{l+1}, j_l) \otimes \one	
& \text{ if } \card{S} > 1
\\
\one^{\otimes n+1} 
& \text{ if } \card{S} \leq 1
\, ,
\end{cases}
\end{eqnarray*}
and with $\CNOT_n(c, t) \in \unOp{\Hil_W}$ we have
$\op{R}_S \in \unOp{\Hil_W \times \Hil_A}$.
Moreover, we set 
\begin{equation*}
\op{V}_S
=
\begin{cases}
\CNOT_{n+1}(j_1 + 1, 0)
& \text{ if } \card{S} \geq 1
\\
\one
& \text{ if } S = \emptyset
\, ,
\end{cases}
\end{equation*}
where $\CNOT_{n+1}(j_1 + 1, 0)$ is the controlled-NOT 
in $\unOp{\Hil_W \times \Hil_A}$
with the control qubit $j_1 +1$ 
with $j_1$ being the smallest element in $S$ and the target qubit being the qubit 
in the ancilla register $\Hil_A$.

\begin{figure}[H]
\centering
\includegraphics{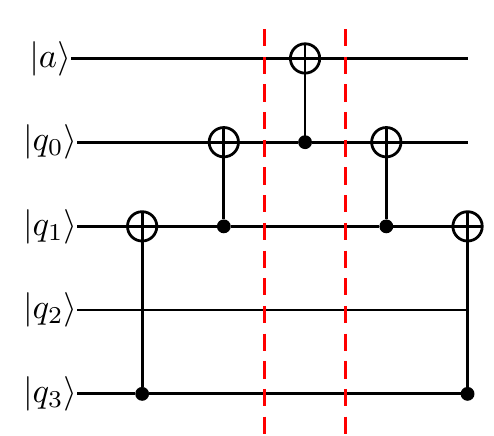}
\caption{Illustration of the circuit for $\op{U}_{\hat{S}\bitscal}$ 
for $S = \{0, 1, 3\} \in P[4]$}
\label{fig:exofUshat}
\end{figure}

Acting on computational basis vectors $\ket{x} \otimes \ket{a} \in \Hil_W \otimes \Hil_A$
the operator $\op{R}_{S}$ gives
\begin{equation*}
\op{R}_{S} \ket{x} \otimes \ket{a}
= 
\ket{y} \otimes \ket{a}
\, ,
\end{equation*}
where
\begin{equation*}
y_j
=
\begin{cases}
x_j		& \text{ if } j \not\in S
\\
x_{j_{l-1}} \bitplus \dotsm \bitplus x_{j_{\betrag{S}}} & \text{ if } j = j_l \in S \setminus \{ j_1 \}
\\
\hat{S} \bitscal x	& \text{ if } j = j_1 \in S
\, ,
\end{cases}
\end{equation*}
and in the last line we have used that 
$x_{j_{1}} \bitplus \dotsm \bitplus x_{j_{\betrag{S}}} = \hat{S} \bitscal x$.
Acting with $\op{V}_S$ on a computational basis vector
$\ket{y} \otimes \ket{a} \in \Hil_W \otimes \Hil_A$
we obtain
\begin{equation*}
\op{V}_S \ket{y} \otimes \ket{a}
=
\begin{cases}
\ket{y} \otimes \ket{y_{j_1} \bitplus a}
& \text{ if } \card{S} \geq 1
\\
\ket{y} \otimes \ket{a}
& \text{ if } S = \emptyset
\, .
\end{cases}
\end{equation*}
Since $\op{V}_S$ writes the binary sum of the ancilla bit and $y_{j_{1}}$ into the 
ancilla register and leaves the work register unchanged, it follows that
\begin{equation*}
\op{V}_S\op{R}_{S} \ket{x} \otimes \ket{a}
= 
\ket{y} \otimes\ket{a \bitplus \hat{S} \bitscal x} 
\, ,
\end{equation*}
and thus
\begin{equation*}
\op{R}^{\dagger}_{S}\op{V}_S\op{R}_{S} \ket{x} \otimes \ket{a}
= 
\ket{x} \otimes \ket{a \bitplus \hat{S} \bitscal x} 
\, ,
\end{equation*}
that is, $\op{R}^{\dagger}_{S}\op{V}_S\op{R}_{S}$ acts exactly as we
require $\op{U}_{\hat{S}\bitscal}$ to act in \eqref{equa:OracleOp.ImplU_S.20}.
Consequently, we can construct the circuit for $\op{U}_{\hat{S}\bitscal}$ as
\begin{equation*}
\op{U}_{\hat{S}\bitscal}
=
\op{R}^\dagger_S \op{V}_S \op{R}_S.
\end{equation*}
Figure \ref{fig:exofUshat} shows $\op{U}_{\hat{S}\bitscal}$ for the case
$n=4$ and $S = \{0, 1, 3\} \in P[4]$.

In order to construct $\op{U}_S$ with the help of $\op{U}_{\hat{S}\bitscal}$,
we will also use the controlled phase multiplication operators
\begin{equation*}
\begin{aligned}
\text{CM}_0(\alpha) 
& =  \one^{\otimes n} \otimes \bigl( \E^{\I \alpha} \ket{0}\bra{0} + \ket{1}\bra{1} \bigr)  
\\ 
\text{CM}_1(\alpha)
& = \one^{\otimes n} \otimes \bigl( \ket{0}\bra{0} +  \E^{-\I \alpha} \ket{1}\bra{1} \bigr) 
\, .
\end{aligned}
\end{equation*}
The action of the controlled phase multiplication operators 
$\text{CM}_0(\alpha)$ and $\text{CM}_1(\alpha)$ on a basis vector
$\ket{x} \otimes \ket{a}$ in $\Hil_W \otimes \Hil_A$ is as follows
\begin{equation*}
\begin{aligned}
\text{CM}_0(\alpha) \ket{x} \otimes \ket{a}
& = 
\E^{\I\alpha (1-a)} \ket{x} \otimes \ket{a} 
\\
\text{CM}_1(\alpha) \ket{x} \otimes \ket{a}
& = 
\E^{-\I \alpha a} \ket{x} \otimes \ket{a}
\, ,
\end{aligned}
\end{equation*}
which implies
\begin{equation*}
\text{CM}_1(\alpha)\text{CM}_0(\alpha) \ket{x} \otimes \ket{a}
=
\E^{\I \alpha (1-2a)} \ket{x} \otimes \ket{a}
\, .
\end{equation*}
We can write $\text{CM}_0(\alpha)$ and $\text{CM}_1(\alpha)$ as
\begin{equation*}
\begin{aligned}
\text{CM}_0(\alpha)
& = 
\one^{\otimes n} \otimes \op{X} \op{P}(\alpha) \op{X} 
\\ 
\text{CM}_1(\alpha)
& =
\one^{\otimes n} \otimes \op{P}(-\alpha) 
\, ,
\end{aligned}
\end{equation*}
where $\op{X}$ is the NOT gate (or bit-flip gate) and 
$\op{P}(\alpha) = \ket{0}\bra{0} + \E^{\I\alpha} \ket{1} \bra{1}$
is the phase-shift gate. This follows from
\begin{eqnarray*}
\text{CM}_0(\alpha) 
& = & \one^{\otimes n} \otimes
\bigl( \E^{\I\alpha} \ket{0}\bra{0} + \ket{1}\bra{1} \bigr) 
\\
& = & \one^{\otimes n} \otimes 
\op{X} \bigl(\E^{\I\alpha} \ket{1}\bra{1} + \ket{0}\bra{0}  \bigr) \op{X} 
\\ 
& = & 
\one^{\otimes n} \otimes
\op{X} \op{P}(\alpha) \op{X} 
\end{eqnarray*}
and similarly for $\text{CM}_1(\alpha)$
\begin{equation*}
\text{CM}_1(\alpha) 
=
\one^{\otimes n} \otimes 
\bigl(  \ket{0}\bra{0} + \E^{-\I \alpha} \ket{1}\bra{1} \bigr) 
= 
\one^{\otimes n} \otimes \op{P}(-\alpha)
\, .
\end{equation*}
For each $S \subset \{0, 1, \ldots, n-1\}$ we construct the circuit
\begin{equation*}
\op{U}_S(\alpha)
=\
\op{U}_{\hat{S} \bitscal } \text{CM}_1(\alpha) \text{CM}_0(\alpha) \op{U}_{\hat{S} \bitscal }
\, .
\end{equation*}
Applying $\op{U}_S(\alpha)$ to $\ket{x} \otimes \ket{a}$ gives
\begin{eqnarray*}
\op{U}_S(\alpha) \ket{x} \otimes \ket{a}
& = &
\op{U}_{\hat{S} \bitscal} \text{CM}_1(\alpha) \text{CM}_0(\alpha) \op{U}_{\hat{S} \bitscal} 
\ket{x} \otimes \ket{a}
\\ \nonumber
& = & 
\op{U}_{\hat{S} \bitscal} \text{CM}_1(\alpha) \text{CM}_0(\alpha) 
\ket{x} \otimes \ket{a \bitplus (\hat{S} \bitscal x)}  
\\ \nonumber
& = & 
\op{U}_{\hat{S} \bitscal} \E^{\I\alpha(1 - 2(a \bitplus (\hat{S} \bitscal x)))} 
\ket{x} \otimes \ket{a \bitplus (\hat{S} \bitscal x)} 
\\ \nonumber
& = &
\E^{\I\alpha(1 - 2(a \bitplus (\hat{S} \bitscal x)))}  \ket{x} \otimes \ket{a} 
\\ 
& = &
\begin{cases}
	\E^{\I\alpha\chi_S(x)}		\ket{x} \otimes \ket{0}	& \text{ if } a = 0
	\\
	\E^{-\I \alpha\chi_S(x)}	\ket{x} \otimes \ket{1}	& \text{ if } a = 1
	\, ,
\end{cases}
\end{eqnarray*}
where we have used \eqref{equa:OracleOp.ImplU_S.10} in the last equation.
Consequently, 
\begin{equation*}
\op{U}_{S}(\alpha)
=
\sum_{x \in \bcube^n} 
\ket{x}\bra{x} \otimes \bigl(
\E^{\I\alpha\chi_S(x)} \ket{0}\bra{0}
+
\E^{-\I\alpha\chi_S(x)} \ket{1}\bra{1} 
\bigr) 
\, ,
\end{equation*}
which is of the form \eqref{equa::OracleOp.FourierCoeff.90} and together with
\eqref{equa::OracleOp.FourierCoeff.110} completes the construction of $\op{U}_f$.

The operators $\op{R}_S, \op{U}_{\hat{S}\bitscal}, \op{U}_S$ and 
$\op{U}_f$ have been implemented in
Qiskit~\cite{soft:Qiskit}
and are available on GitHub 
(see Section Code Availability at the end).
%
\section{Application: Binary Optimization}
\label{sect:BinOpt}
\subsection{Quadratic Binary Optimization}
\label{subsect:BinOpt.QuadBO}
We demonstrate the potential of the Non-Boolean oracle operator
in solving binary optimization problems by first considering
a quadratic unconstrained binary optimization (QUBO) problem.
The typical approach to solve this kind of problem
is by adiabatic quantum computing~\cite{book:Scherer2019}.

Let the function $B: \bcube^n \to \reellZ$, for which we want to find the
extrema, be given by
\begin{equation*}
B(x) = \sum_{i,j = 0}^{n-1} Q_{ij} x_i x_j
\, ,
\end{equation*}
where $Q \in \Mat( n \times n, \reellZ)$ is a symmetric matrix.
From \eqref{equa::OracleOp.FourierCoeff.55} we infer
\begin{equation*}
x_i^2 = x_i = \frac{1}{2} (1- \chi_{\{i\}}(x)) = \frac{1}{2} (\chi_{\emptyset}(x) - \chi_{\{i\}}(x))
\end{equation*}
and for $i \neq j$
\begin{eqnarray*}
x_ix_j 
& = &
\frac{1}{4}(\chi_{\{i, j\}}(x) - 1 + 2x_i + 2x_j)
\\
& = &
\frac{1}{4}(\chi_{\{i, j\}}(x) + \chi_{\emptyset}(x) - \chi_{\{i\}}(x) - \chi_{\{j\}}(x))
\, .
\end{eqnarray*}
Hence,
\begin{eqnarray*}
B(x) 
& = &
\sum_{i = 0}^{n - 1}Q_{ii}x_i^2 + 2\sum_{i=0}^{n-1} \sum_{j=i+1}^{n-1} Q_{ij}x_ix_j
\\
& = &
\frac{1}{2}\sum_{i = 0}^{n - 1}Q_{ii} (\chi_{\emptyset}(x) - \chi_{\{i\}}(x)) 
\\
& &
+ \frac{1}{2} \sum_{i=0}^{n-1} \sum_{j=i+1}^{n-1}Q_{ij} (\chi_{\{i, j\}}(x) + \chi_{\emptyset}(x) - \chi_{\{i\}}(x) - \chi_{\{j\}}(x)) 
\\
& = &
\sum_{S\in P[n]} \hat{B}(S)\chi_S(x)
\, ,
\end{eqnarray*}
and thus the Fourier coefficients of $B(x)$ can be read of as functions of the $Q_{ij}$ as
\begin{eqnarray*}
	\hat{B}(\emptyset) &=& \frac{1}{2} \Big( \sum_{i=0}^{n-1} Q_{ii} + \sum_{i=0}^{n-1} \sum_{j=i+1}^{n-1} Q_{ij} \Big)
	\\
	\hat{B}(\{i\}) &=& - \frac{1}{2} \sum_{j=0}^{n-1} Q_{ij} 
	\\
	\hat{B}(\{i, j\}) &=& \frac{1}{2} Q_{ij}
	\fadist \text{for } i \neq j
	\\ 
	\hat{B}(S) & = & 0
	\fadist \forall S \in P[n] : \card{S} > 2
	\, .
\end{eqnarray*}
With the matrix $Q$ we also define
\begin{eqnarray*}
q_{\pm} 
& = &
\sum_{\substack{i,j = 0 \\ Q_{ij} \gtrless}}^{n-1} Q_{ij} 
\gtreqless 0
\\
\norm{Q}_{1,1} 
& = &
\sum_{i,j = 0}^{n-1} \betrag{ Q_{ij} } 
=
q_{+} - q_{-}
\, .
\end{eqnarray*}
Since $x \in \bcube^n$, we have $q_{-} \leq B(x) \leq q_{+}$ and thus
\begin{equation*}
b_{\pm}(x) := \pm \frac{q_{\pm} - B(x)}{\norm{Q}_{1,1}} \frac{\pi}{2}
\in [0, \frac{\pi}{4}]
\, .
\end{equation*}
From \eqref{equa:AA.ExtremNBAA.110} we know that the set of $x$
for which $b_{-}$ takes its maximum value is the same set where $B$
takes its maximum value and the set of $x$ where $b_{+}$ takes its
maximum value is the set where $B$ takes its minimum.

\begin{figure}[h!]
\centering
\includegraphics[scale=0.32]{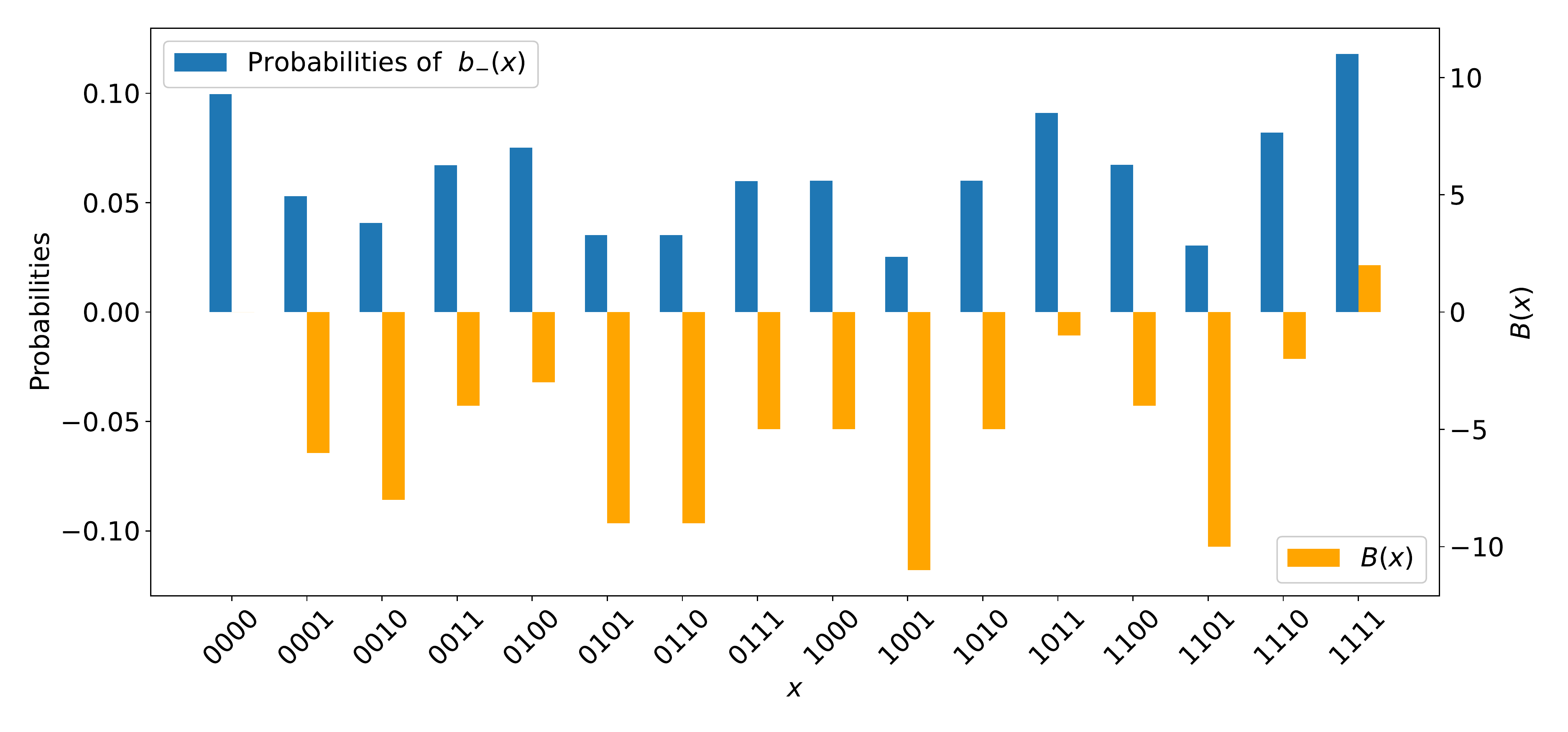}
\caption{The blue bars and the left scale show the 
probability to find $x \in \bcube^4$ after running the Non-Boolean
amplitude amplification for $b_{-}$ with $\theta_{-} = 0.296, \tilde{K}_{-}= 5$ and
$\lambda_{\tilde{K}_{-}}(\theta_{-}) = 22.83$ in Qiskit. 
The orange bars and the right scale show the values of $B(x)$.
The biggest probability is
seen to be at $x=1111$, which is the value where $B$ takes its maximum.
The second highest probability is at  the point $x=0000$, where $B$
takes the next highest value.}
\label{fig:BinOpt.QuadBO.b_minus}
\end{figure}

Hence, our aim will be to find the maxima for $b_{\pm}$.
As shown in Section~\ref{subsect:AA.ExtremNBAA}, we can
accomplish this by using the Non-Boolean amplitude amplification 
algorithm with the oracle operator $\op{U}_{b_{\pm}}$ to amplify the 
probabilities of $x$ for lower values of $\cos\bigl(b_{\pm}(x)\bigr)$ 
and identify the maximum or minimum of $B(x)$. For this, we can calculate
the Fourier coefficients of the $b_{\pm}$ directly from the Fourier coefficients 
of $B$ as
\begin{eqnarray*}
\hat{b}_{\pm}(\emptyset) 
& = & 
\pm \frac{\pi}{2} \frac{q_{\pm} - \hat{B}(\emptyset)}{\norm{Q}_{1,1}}
\\
\hat{b}_{\pm}(S) 
& = & 
\mp \frac{\pi}{2\norm{Q}_{1,1}} \hat{B}(S)
\fadist \forall S \in P[n] : \card{S} > 0
\, .
\end{eqnarray*}
As a simple example, we consider the case $n= 4$ and 
the function used in \cite{art:Glover-et-al2019}
for illustration
\begin{equation*}
B(x)
=
-5 x_3 - 3 x_2 - 8 x_1 - 6 x_0 + 4 x_3 x_2 + 8 x_3 x_1 + 2 x_2 x_1 + 10 x_2 x_0
\, ,
\end{equation*}
which can be written (using the usual binary ordering) as
\begin{equation*}
B(x) =
\begin{pmatrix}
x_3, x_2, x_1, x_0
\end{pmatrix}
\underbrace{
\begin{pmatrix}
-5	&	2	&	4	&	0	\\ 
2	&	-3	&	1	&	0 	\\ 
4	&	1	&	-8	&	5	\\ 
0	&	0	&	5	&	-6 
\end{pmatrix}}_{=Q}
\begin{pmatrix}
x_3 \\ x_2 \\ x_1 \\ x_0
\end{pmatrix}
\, .
\end{equation*}
This function has the extrema
\begin{equation*}
\begin{aligned}
\min_{x \in \bcube^4} \{ B(x) \} 		
& = & 
-11
& =
B(1001)
\\
\max_{x \in \bcube^4} \{ B(x) \} 		
& = & 
2
& =
B(1111)
\, .
\end{aligned}
\end{equation*}
We simulate the algorithms for $b_{\pm}$ in Qiskit~\cite{soft:Qiskit}.
The value $x_{-, \max} = 1111$ is where $b_{-}(x)$ is maximal,
$\cos\bigl(b_{-}(x)\bigr)$ is minimal
and for which the probability will be amplified the most, 
when the amplitude amplification is applied to $b_{-}(x)$,
as can be seen in 
Figure~\ref{fig:BinOpt.QuadBO.b_minus}.

\begin{figure}[h!]
\centering
\includegraphics[scale=0.32]{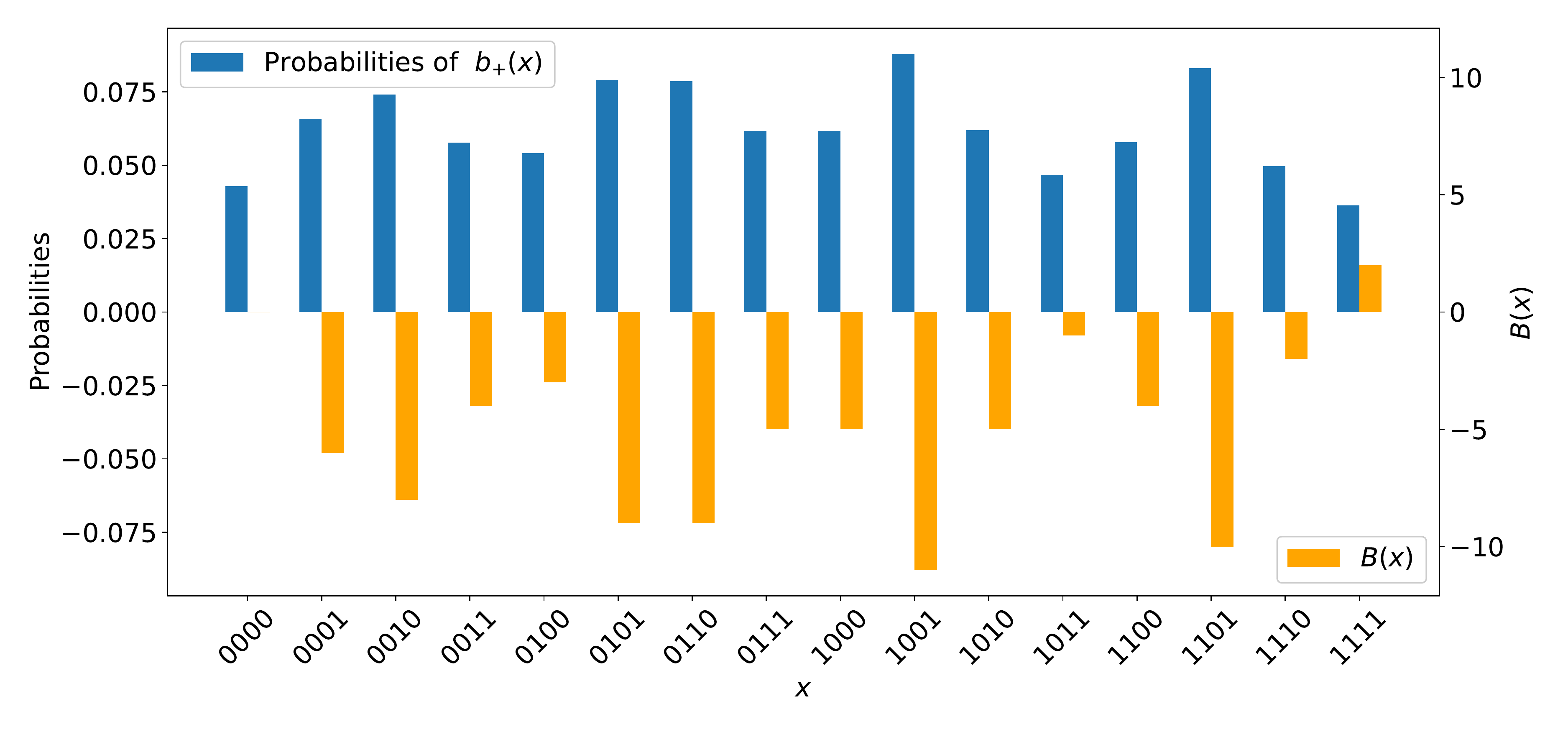}
\caption{The blue bars and the left side show the
probability to find $x \in \bcube^4$ after running the Non-Boolean
amplitude amplification for $b_{+}$ with $\theta_{+} = 0.499, \tilde{K}_{+}= 3$
and $\lambda_{\tilde{K}_{+}}(\theta_{+}) = 7.95$ in Qiskit. 
The biggest probability is
seen to be at $x=1001$, which is the value where $B$ takes its minimum.}
\label{fig:BinOpt.QuadBO.b_plus}
\end{figure}

On the other hand,  $x_{+, \max} = 1001$ is the value where $\cos\bigl(b_{+}(x)\bigr)$ 
is minimal and for which the probability will be amplified the most, 
when the amplitude amplification is applied to $b_{+}(x)$,
as shown in 
Figure~\ref{fig:BinOpt.QuadBO.b_plus}.

Figures \ref{fig:BinOpt.QuadBO.b_minus} and \ref{fig:BinOpt.QuadBO.b_plus}
also show that not only are the extrema amplified the most, 
but the probabilities for all other points $x$ are amplified
in accordance with the size of the values $B(x)$.
This means that even if a read-out after a simulation will not
yield the actual extremal point, it will most likely yield a
point with a value close to the extremum. This is
particularly useful in real world optimization problems,
such as the Travelling Salesman, where it might not be 
crucial to find the true extremum, but may be good enough to find
something near to it.
\subsection{Higher Order Binary Optimization}
\label{subsect:BinOpt.HighOrdOpt}
In more general binary optimization problems of order $m \geq 1$, 
the function, for which we want to find the extrema, is of the form 
\begin{equation*}
F(x) 
=
\sum_{{i_1,\ldots,i_m = 0}}^{n - 1} Q_{i_1,\ldots,i_m} x_{i_1} \ldots x_{i_m} 
\, .
\end{equation*}
For the sets $\{ i_1, \dotsc, i_l \}$ with $l \leq m$ and $i_j < i_{j+1}$ we have
\begin{eqnarray*}
\chi_{\{i_1, \dotsc, i_l\}}(x) 
& = & 
(-1)^{x_{i_1} + \dotsm + x_{i_l}}
\\ 
& = & 
(1 - 2x_{i_1})\dotsm(1 - 2x_{i_l}) 
\\
& = & 
1 - 2x_{i_1} - \dotsm - 2x_{i_l} + 4x_{i_1}x_{i_2} + \dotsm + (-2)^l x_{i_1} \dotsm x_{i_l} 
 \, ,
\end{eqnarray*}
such that, as in the quadratic case, the products $x_{i_1} \dotsc x_{i_m}$ can be
written as linear combinations of parity functions $\chi_S(x)$ for sets with cardinality
not bigger than $m$, that is, 
\begin{equation*}
x_{i_1} \dotsm x_{i_m} 
=
\sum_{\substack{S \in P[n]: \\ \betrag{S} \leq m}} 
G(S)_{i_1, \dotsc , i_m}\chi_{S}(x)
\, ,
\end{equation*}
where the $G(S)_{i_1, \dotsc , i_m}$ can be easily determined.
Hence,
\begin{eqnarray*}
F(x) 
& = &
\sum_{\substack{S \in P[n]: \\ \betrag{S} \leq m}} 
\sum_{i_1, \dotsc, i_m = 0}^{n - 1} Q_{i_1, \dotsc, i_m} 
G(S)_{i_1, \dotsc , i_m} \chi_{S}(x)
 = 
\sum_{S\in P[n]} \hat{F}(S)\chi_S(x)
\, ,
\end{eqnarray*}
and the Fourier coefficients of $F(x)$ can again be read off as
\begin{equation*}
\hat{F}(S)
=
\begin{cases}
\sum_{i_1, \dotsc, i_m = 0}^{n - 1} Q_{i_1, \dotsc, i_m} 
G(S)_{i_1, \dotsc , i_m} & \text{ if } \betrag{S} \leq m
\\
0 & \text{ if } \betrag{S} > m
\, .
\end{cases}
\end{equation*}
This also shows that the number of non-zero Fourier coefficients in a binary
optimization problem of order $m$ is $\Landau{n^m}$, and it follows that for
$m$ not too large, our method of constructing $\op{U}_f$ remains efficient.
\section{Conclusion}
\label{sect:Conc}
We have presented a circuit to implement the oracle operator 
\begin{equation*}
\op{U}_f
=
\sum_{x \in \bcube^n} 
\ket{x}\bra{x} \otimes  
\bigl(\E^{\I f(x)} \ket{0}\bra{0}   + \E^{-\I f(x)}\ket{1}\bra{1} \bigr) 
\end{equation*}
for any function $f: \bcube^n \to \reellZ$. The circuit in this
implementation only uses the Pauli-$\op{X}$ and the phase-shift gates.

While such an implementation is a valuable
construction in its own right, 
for example, in implementations of quantum Fourier
transformations on finite Abelian groups~\cite{book:Scherer2019},
it is particularly useful in the 
context of the Non-Boolean amplitude amplification algorithm,
where it can be utilized to attempt to solve binary optimization problems
with a non-adiabatic algorithm.

It thus suggests the construction of a non-adiabatic algorithm for many real world problems, 
such as Partitioning Problems, Binary Integer Linear Programming, 
Covering and Satisfiability Problems, Coloring Problems, 
Hamiltonian Cycles, Tree Problems and Graph Isomorphisms~\cite{art:Lucas2014}.
For such problems the Fourier coefficients required for the construction
of the respective oracle operators
can be efficiently calculated from the problem parameters without
evaluating the function for which we want to find the extrema.

We will explore the use of Non-Boolean amplitude amplification in solving
such binary optimization problems in a non-adiabatic algorithm
in more detail in a forthcoming paper.
%
\section*{Declarations}
\subsection*{Code Availability}
\label{sect:CodeAv}
The code of this study is partly based on code from Shyamsundar and is
openly available as a Python Jupyter notebook at the following URL: 
\href{https://gitlab.com/fatemamelg/oracle-operators-for-non-boolean-functions}{https://gitlab.com/fatemamelg/oracle-operators-for-non-boolean-functions} 
under the directory named "Oracle Operators for Non-Boolean Functions".
%
\bibliographystyle{unsrturl}

%
\end{document}